**Deep Learning-based Radiomic Features for Improving Neoadjuvant Chemoradiation Response Prediction in Locally Advanced Rectal Cancer**

Jie Fu[1]*, Xinran Zhong[2]*, Ning Li[3], Ritchell Van Dams[1], John Lewis[4], Kyunghyun Sung[2], Ann C. Raldow[1], Jing Jin[3], X. Sharon Qi[1]

1 University of California, Los Angeles, Department of Radiation Oncology, Los Angeles, CA, USA, 90095

2 University of California, Los Angeles, Department of Radiological Sciences, Los Angeles, CA, USA, 90095

3 Department of Radiation Oncology, Cancer Hospital Chinese Academy of Medical Sciences，Beijing, China

4 Cedars Sinai Medical Center, Los Angeles, CA

* Authors contributed equally in this study

Email: jiefu@mednet.ucla.edu and xqi@mednet.ucla.edu

Abstract

**Abstract**

Radiomic features achieve promising results in cancer diagnosis, treatment response prediction, and survival prediction. Our goal is to compare the handcrafted (explicitly designed) and deep learning (DL)-based radiomic features extracted from pre-treatment diffusion-weighted magnetic resonance images (DWIs) for predicting neoadjuvant chemoradiation treatment (nCRT) response in patients with locally advanced rectal cancer (LARC). 43 patients receiving nCRT were included. All patients underwent DWIs before nCRT and total mesorectal excision surgery 6-12 weeks after completion of nCRT. Gross tumor volume (GTV) contours were drawn by an experienced radiation oncologist on DWIs. The patient-cohort was split into the responder group (n=22) and the non-responder group (n=21) based on the post-nCRT response assessed by postoperative pathology, MRI or colonoscopy. Handcrafted and DL-based features were extracted from the apparent diffusion coefficient (ADC) map of the DWI using conventional computer-aided diagnosis methods and a pre-trained convolution neural network, respectively. Least absolute shrinkage and selection operator (LASSO)-logistic regression models were constructed using extracted features for predicting treatment response. The model performance was evaluated with repeated 20 times stratified 4-fold cross-validation using receiver operating characteristic (ROC) curves and compared using the corrected resampled t-test. The model built with handcrafted features achieved the mean area under the ROC curve (AUC) of 0.64, while the one built with DL-based features yielded the mean AUC of 0.73. The corrected resampled t-test on AUC showed P-value < 0.05. DL-based features extracted from pre-treatment DWIs achieved significantly better classification performance compared with handcrafted features for predicting nCRT response in patients with LARC.

**Keywords:** Deep learning, Radiomics, Treatment response prediction

## 1. Introduction

Colorectal cancer is the third most common cancer diagnosed and the second most common cause of cancer deaths in the US [1]. Rectal cancer accounts for about 30% of all colorectal cancer diagnoses [1]. Treatment for rectal cancer is based largely on the stage at diagnosis. Locally advanced rectal cancer (LARC) is commonly treated with neoadjuvant chemoradiation therapy (nCRT) followed by total mesorectal excision (TME) and adjuvant chemotherapy[2,3]. Tumor response to nCRT is associated with recurrence and survival and can serve as a prognostic factor [4,5]. 15-27% of patients who undergo such treatment achieve pathologic complete response (pCR) [6]. TME is a highly invasive procedure with the potential risk of morbidity and functional complications. Achieving early prediction of tumor response using noninvasive approaches may allow for individualized patient management and potential avoidance of TME following nCRT.

Magnetic resonance imaging (MRI) is widely used in rectal cancer diagnosis and staging as it provides excellent soft tissue contrast for tissue characterization. Specifically, increasing evidence has shown that diffusion-weighted images (DWIs), providing tissue cellularity information, aids the assessment of rectal cancer response to neoadjuvant treatment [7]. DWI is recommended to be routinely acquired in clinical guidelines [8]. The interpretation of DWI has gradually shifted from qualitative evaluation to quantitative assessment. For example, the apparent diffusion coefficient (ADC) map was one major quantitative map calculated from DWI. However, several studies showed that the mean pretreatment tumor ADC value was not a reliable indicator for predicting treatment response [9,10].

Radiomics is an emerging field of studies where a large number of medical image features are extracted in order to achieve better clinical diagnosis or decision support [11]. The conventional radiomics analysis typically involves extraction and analyzing quantitative imaging features from the previously defined region of interests (ROI) on one or multiple image modalities with the ultimate goal to obtain predictive or prognostic models. Previous studies showed that handcrafted, or explicitly designed, features extracted from the ADC ROI have predictive power for early nCRT treatment response in LARC patients [12,13]. However, handcrafted features are lower-order image features and limited to current expert knowledge. Another type of radiomic feature is deep learning (DL)-based extracted from the pre-trained convolutional neural networks (CNN) via transfer learning [14,15]. Several studies have demonstrated that the DL-based features showed promising performance in breast cancer diagnosis, ovarian cancer recurrence prediction, and glioblastoma multiforme survival prediction [16–18]. To our knowledge, no published study has investigated the DL-based features for managing LARC patients.

In this work, we first aimed to construct radiomics classifiers based on the handcrafted and DL-based radiomic features extracted from pre-treatment DWIs. Then, we compared the performance of the two classifiers to predict post-nCRT response in patients with LARC.

## 2. Materials and methods
### 2.1. Dataset

We identified forty-three consecutive patients with locally advanced rectal cancer (LARC) treated from December 2015 to December 2016 at a single institution. All patients received concurrent capecitabine with a total prescription dose of 50 Gy in 25 fractions, followed by the TME surgery after 6-12 weeks of the nCRT completion. The resection specimens were evaluated by an expert pathologist. Patients were separated into good responders (GR) and non-GR groups based on the postoperative pathology report, MRI or colonoscopy. The GR group consisted of patients with either complete response (evaluated by pathology or MRI and colonoscopy) or partial response (assessed by pathology), and the non-GR group consisted of patients with stable disease (assessed by pathology) and progressive disease (confirmed by CT/MR).

All patients underwent pre-treatment DWIs before the nCRT. The DWI images were acquired using single-shot echo planar imaging (ssEPI) sequence on two 3-Tesla MR scanners. MR imaging parameters are summarized in Table 1. For each patient, the ADC map was computed using the equation $ADC = -\frac{1}{800}\ln(\frac{S}{S_0})$, where $S_0$ and S correspond to MR voxel intensities at b-values of 0 s/mm$^2$ and 800 s/mm$^2$. Gross tumor volume (GTV) of the primary tumor was delineated by an experienced board-certified oncologist.

| Scanner model | Patient number | TR/TE (ms) | Spatial resolution | Field of view (cm$^2$) | Slice thickness (mm) | b value (s/mm$^2$) |
|---|---|---|---|---|---|---|
| Discovery MR750 | 36 | 2600/74 | 256×256 | 38$^2$ or 40$^2$ | 5 | 0,500,800,1000 |
| Signa HDxt | 7 | 4500-6000/64-67 | 256×256 | 32$^2$ - 40$^2$ | 5 or 6 | 0,800 |

Table 1. MR imaging parameters

## 2.2. Feature extraction

2.1.1. Handcrafted features

105 handcrafted features were extracted from the ADC map within the GTV contour for each patient using PyRadiomics package (version 2.1.2) [19]. Extracted features consisted of 14 shape-based features, 18 first-order statistic features, and 73 textural (second-order statistic) features. The methods used for extracting textural feature were gray level co-occurrence matrix, gray-level size zone matrix, gray level run length matrix, gray level dependence matrix, and neighborhood gray-tone difference matrix. Shape-based features describe the shape characteristics of the GTV contour. First-order statistic features describe the distribution of voxel intensities within the GTV contour. Textural features describe the patterns or second-order spatial distributions of the voxel intensities.

2.2.2. DL-based features

The publicly-available pre-trained CNN, VGG19 [20], was used to extract DL-based features. The network was trained using approximately 1.2 million images from the ImageNet database [21] for classifying nature images into 1000 objects. Figure 1 shows the network architecture. It contained

16 convolutional layers followed by 3 fully-connected layers. 5 max-pooling layers were inserted across convolutional and fully-connected layers to reduce model parameter number for controlling overfitting and help achieve partial invariance to small translations. For each patient, a square region of interest (ROI) was selected around the tumor in the transverse slice having the largest tumor area. The ROI size was set based on the maximum tumor dimension. The ADC ROI was extrapolated to 224 by 224 for matching the original VGG19 design and then input into the pre-trained model for feature extraction. We adopted the feature extraction method proposed by Antropova *et al.* [16]. As shown in Figure 1, five DL-based feature vectors were extracted by average-pooling the feature maps after max-pooling layers. Each feature vector was normalized with its Euclidean norm and then concatenated to one feature vector, which was normalized again to acquire the feature vector consisting of 1472 features. After extracting features for all patients, a cutoff on feature variance was used to select 105 DL-based features with the highest variance.

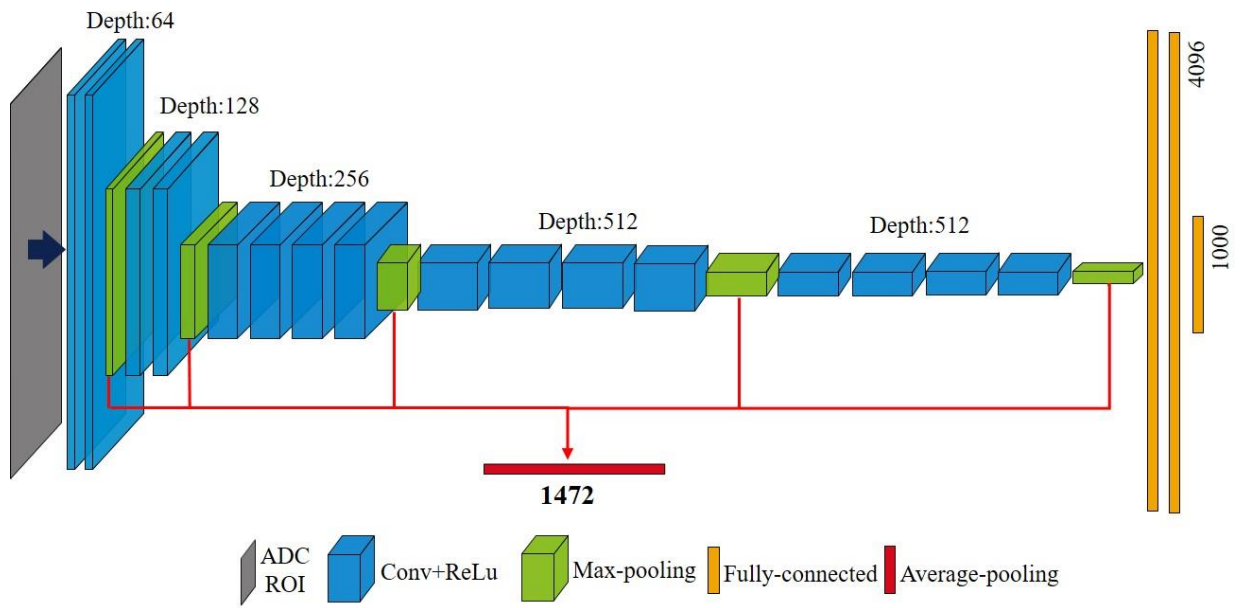

Figure 1. VGG19 architecture and feature extraction scheme. It contained 5 max-pooling layers inserting across 16 convolutional layers and 3 fully-connected layers. Feature maps and feature vectors, following each layer, are shown as cuboids and rectangles, respectively. The feature map depth and feature number are shown. For feature extraction, the network took an ADC ROI as input. 1472 DL-based features were extracted from max-pooling feature maps by average-pooling along the spatial dimensions. Conv, convolutional layer; ReLu, rectified linear unit.

### 2.3. Classification and evaluation

The least absolute shrinkage and selection operator (LASSO) penalized logistic regression [22,23] was used for classification using radiomic features (Python version 2.7.13). The LASSO regularization was selected to handle the high feature dimension. The handcrafted classifier and DL-based classifier were trained using handcrafted features and DL-based features, respectively. Regularization parameter was optimized by grid searching with repeated 20 times stratified 4-fold cross-validation. For each cross-validation, stratified random sampling was used to split the patient cohort was into 4 folds, where 3 folds were used as the training set to train the classifier and the remaining one as the testing set for evaluation.

The performances of the handcrafted and DL-based classifiers were evaluated using the average area under the receiver operating curve (AUC) of 20 cross-validation repetitions. The corrected resampled t-test [24,25] was conducted to compare the AUC results for two classifiers. P-value <0.05 was considered to indicate a significant difference.

## 3. Results
### 3.1. Patient characteristics

Table 2 summarizes the clinical characteristics of our patient cohort. 22 (51.2%) patients achieved GR after nCRT. Among the 22 GR patients, there were 14 (63.6%) men and 8 (36.4%) women. Among 21 non-GR patients, there were 14 (66.7%) men and 7 (33.3%) women.

| **Characteristic** | **GR (n=22)** | **nGR (n=21)** | **Total (n=43)** |
|---|---|---|---|
| Gender (male/female) | 14/8 | 14/7 | 28/15 |
| Age (mean, SD, in years) | 53.7 (9.1) | 54.9 (10.9) | 54.3 (10.3) |
| Pre-nCRT TNM staging | | | |
| T stage (2/3/4) | 1/18/3 | 1/16/4 | 2/34/7 |
| N stage (0/1/2) | 5/11/6 | 0/9/12 | 5/20/18 |

Table 2. Patient clinical characteristics; GR, good responder, nGR, non-good responder, SD, standard deviation.

Figure 2 shows the transverse slices of DWIs and ADC maps for the representative GR and non-GR patients. Both patients are male with rectal cancer at the same clinical stage of T3N1. No significant visual differences were observed.

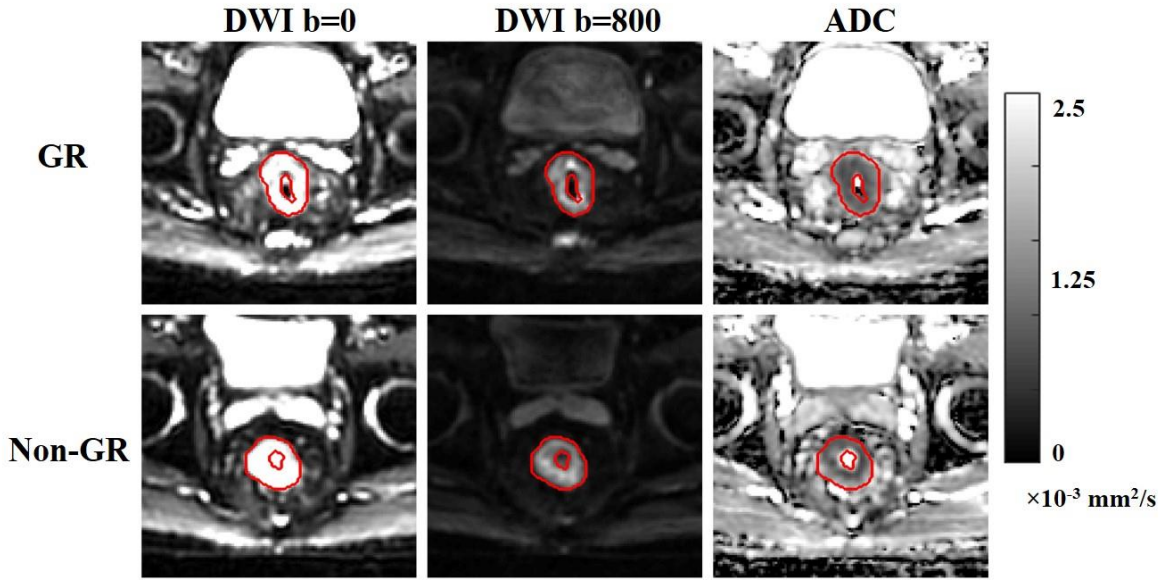

Figure 2. Comparison of the DWI (b=0,800 s/mm²) slice and the ADC slice for the representative GR and non-GR patients. The GTV contours are demonstrated in red. The color bar of the ADC slices is shown.

### 3.2. Classification performance

Figure 3 (a) compares the boxplots of the mean AUC results of 20 cross-validation repetitions for two classifiers. Large deviations were observed due to the small sample size. The AUC of a single repetition varies from 0.51 to 0.73 for the handcrafted classifier, and from 0.58 to 0.80 for the DL-based classifier. The average ROC curves of the two classifiers are shown in Figure 3 (b). The handcrafted classifier achieved the mean AUC of 0.64 (standard error [SE], 0.08) using repeated 20 times 4-fold cross-validation, while the DL-based classifier achieved 0.73 (SE, 0.05). The p-value of the corrected resampled t-test was 0.049, suggesting a significant difference in the AUC results for the handcrafted classifier and DL-based classifier.

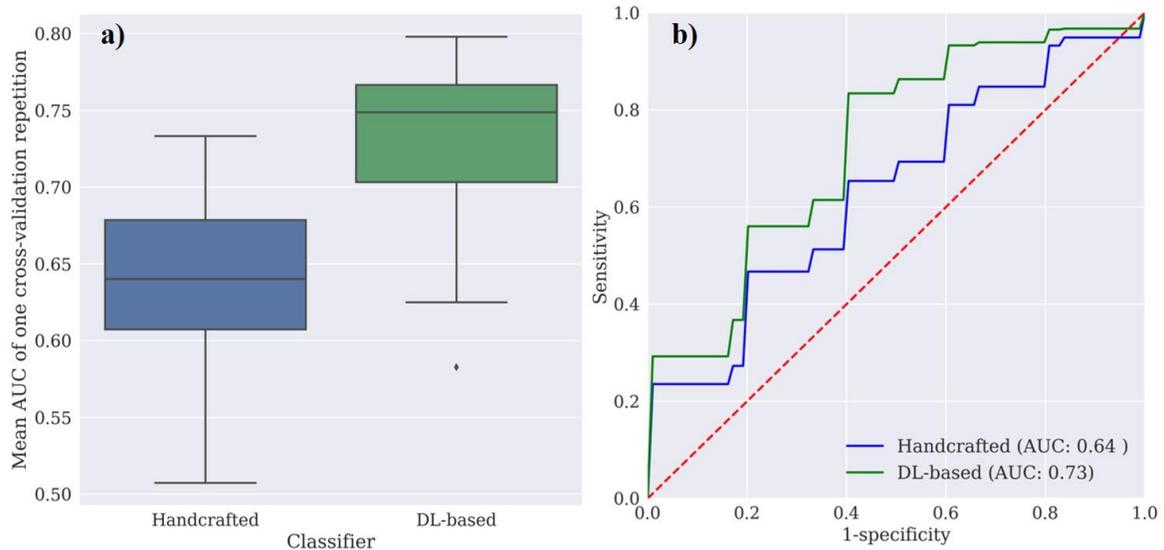

Figure 3. (a) Boxplots of the AUC results of 20 cross-validation repetitions for the handcrafted and DL-based classifiers. The minimum (bottom line), 25$^{th}$ percentile (bottom of the box), median (central line), 75$^{th}$ percentile (top of the box), and maximum (top line) are shown. An outlier is drawn as a diamond sign. (b) The ROC curves for two classifiers in predicting good response versus non-good response using repeated stratified 4-fold cross-validation. AUC results are averaged over 20×4 testing sets.

## 4. Discussion

In this study, we compared the performance of the classifiers built with the handcrafted and DL-based features, extracted from pre-treatment DWI, for predicting the post-nCRT treatment response for a cohort of LARC patients. To our knowledge, this is the first study investigating DL-based features for this application. Compared to the handcrafted features, the DL-based features consisted of more abstract high-level information extracted from DWI images. Our study indicated that the DL-based classifier achieved a significantly better predictive performance than the handcrafted classifier in nCRT response prediction for rectal cancer. Studies showed that the DL-based features achieved better performance in breast cancer diagnosis and glioblastoma survival prediction than the handcrafted features [18,26]. The DL-based features are expected to achieve better performance and more generalizable results in diagnosis, recurrence and survival prediction for other sites as well.

We conducted repeated 4-fold cross-validation for evaluating the model performance as it stabilizes the accuracy estimation [25,27]. The handcrafted classifier achieved the mean AUC of 0.64 for predicting GR vs non-GR, while the DL-based classifier achieved an improved mean AUC of 0.73. Additionally, a fused classifier was constructed by averaging prediction scores of two classifiers. The fused classifier achieved the mean AUC of 0.71, which is better than that for the handcrafted classifier. Nie *et al* [12], using a single run of 4-fold cross-validation, reported the mean AUC of 0.73 for GR and non-GR prediction using DWI handcrafted features on a similar size cohort of 48 patients. The standard error of the mean AUC was not reported. To investigate the cross-validation variation caused by the different data partitions for a small dataset, we conducted 20 independent cross-validation trials using our dataset. It should be noted that 20 independent cross-validation trials are different from the repeated 20 cross-validation since each cross-validation trial has its own optimal hyperparameters, while all 20 cross-validation repetitions need to have the same hyperparameters. The mean AUC of each cross-validation trial ranged from 0.56 to 0.79 for the classifier built with the handcrafted features, and from 0.63 to 0.82 for the one built with the DL-based features. Given a relatively small patient size, a single run of cross-validation may have large bias. Also, different classification models, evaluation protocols, patient number, and response label ratio may result in different prediction accuracy.

We investigated the radiomic features extracted from a single imaging modality of DWI in this study. Several studies showed that including the handcrafted features from T2-weighted MR images and dynamic contrast-enhanced images improved predictive power [12,13]. The DL-based feature extraction scheme can be applied to other MR imaging modalities and may further help improve the prediction accuracy. Comparing the handcrafted and DL-based features extracted from multiparametric MR images for treatment response prediction would be an interesting study to work on in the future.

Our study has several limitations. First, the study sample is small, which may lead to unstable accuracy estimation and suboptimal model performance. A repeated cross-validation method was utilized to reduce the bias in the work. LASSO regularization was implemented to reduce overfitting. A larger training set may result in better model performance. Second, our dataset only contained 9 patients with pathological complete response (pCR). The pCR is defined as the absent of viable tumor cells in the primary and lymph nodes. Small number of pCR patients and unbalanced labels resulted in large standard deviation on the AUCs using either handcrafted features or DL-based features for predicting pCR vs non-pCR. We chose to construct and evaluate the predictive model with the classification labels of GR and non-GR in this preliminary study. A larger dataset is desirable to provide the more reliable estimation for the AUC of pCR and non-pCR prediction. We expect to see better performance from the DL-based features than the handcrafted features in predicting pCR on a larger dataset.

## 5. Conclusion

Our preliminary study showed that the DL-based features extracted from pretreatment DWIs achieved significantly better classification performance for predicting post-nCRT treatment response in LARC patients, in comparison to the handcrafted features. Future work involves validation with a larger dataset and investigating the predictive power of the DL-based features extracted from multiparametric MR images.